# Large Language Models in Biomedical and Health Informatics: A Review with Bibliometric Analysis

Huizi Yu[1]*, Lizhou Fan[1]*, Lingyao Li[1], Jiayan Zhou[2], Zihui Ma[3], Lu Xian[1], Wenyue Hua[4], Sijia He[1], Mingyu Jin[4], Yongfeng Zhang[4], Ashvin Gandhi[5], Xin Ma[6#]

[1] University of Michigan, [2] Stanford University, [3] University of Maryland, [4] Rutgers University, [5] University of California, Los Angeles, [6] Shandong University

* Huizi Yu and Lizhou Fan contributed equally

# Correspondence to maxin@sdu.edu.cn

## Abstract

Large Language Models (LLMs) have rapidly become important tools in Biomedical and Health Informatics (BHI), enabling new ways to analyze data, treat patients, and conduct research. This study aims to provide a comprehensive overview of LLM applications in BHI, highlighting their transformative potential and addressing the associated ethical and practical challenges. We reviewed 1,698 research articles from January 2022 to December 2023, categorizing them by research themes and diagnostic categories. Additionally, we conducted network analysis to map scholarly collaborations and research dynamics. Our findings reveal a substantial increase in the potential applications of LLMs to a variety of BHI tasks, including clinical decision support, patient interaction, and medical document analysis. Notably, LLMs are expected to be instrumental in enhancing the accuracy of diagnostic tools and patient care protocols. The network analysis highlights dense and dynamically evolving collaborations across institutions, underscoring the interdisciplinary nature of LLM research in BHI. A significant trend was the application of LLMs in managing specific disease categories such as mental health and neurological disorders, demonstrating their potential to influence personalized medicine and public health strategies. LLMs hold promising potential to further transform biomedical research and healthcare delivery. While promising, the ethical implications and challenges of model validation call for rigorous scrutiny to optimize their benefits in clinical settings. This survey serves as a resource for stakeholders in healthcare, including researchers, clinicians, and policymakers, to understand the current state and future potential of LLMs in BHI.

**Keywords**: Artificial Intelligence, Biomedical Informatics, Health Informatics, Large Language Models

## 1. Introduction

Large Language Models (LLMs) have emerged as pivotal technologies, redefining the landscape of Natural Language Processing (NLP) and showing significant potential in the intersection of artificial intelligence (AI) and other domains, such as Biomedical and Health Informatics (BHI) [1–3]. The advent of groundbreaking models including OpenAI's Generative Pre-trained Transformer (GPT) [4] has demonstrated its capabilities to process, understand, and generate human-like text by leveraging extensive datasets and sophisticated neural network architectures [5,6]. These advances have set the stage for transformative applications within BHI, a domain where the accuracy and nuance of language understanding significantly impact patient care, medical research, and healthcare delivery.

Since the introduction of models like ChatGPT, the role of LLMs in BHI has been increasingly recognized. These potential applications include clinical decision support, patient engagement enhancement, and medical literature analysis [7–9]. These developments have provided enormous possibilities of not only augmenting traditional methodologies but also paving the way for novel approaches to addressing complex challenges in the healthcare sector.

Our review uniquely contributes to the discourse by offering a comprehensive analysis of LLM applications in BHI in 1,698 papers from January 2022 to December 2023. Through an examination of research themes, scholarly networks, and the evolution of LLM technologies, we delve into the integration and impact of LLMs across various BHI fields. The scope of this study is twofold:

- *Research themes and topics*: We explore the development of LLM algorithms through the lenses of NLP and medical tasks, as well as the LLMs applications in various disease categories, identifying LLM-based applications in BHI.
- *Scholarly networks and partnerships*: Our analysis includes an examination of the collaborative efforts and research networks, underlying the dynamics of research paradigms of LLM research in the BHI domains.

By examining current literature, this review aims to highlight key trends and gaps in current research and further points out the opportunities. Our findings aim to provide a foundation for future research, giving stakeholders important insights to understand and contribute to this rapidly developing field. This review not only shows the enormous prospects of LLMs improving healthcare outcomes but also emphasizes the need to consider ethics and address practical challenges in the case of using LLMs in BHI.

The rest of the paper is organized as follows. We begin by providing the background on the intersection of LLMs and BHI from three perspectives, i.e. the evolution of LLMs, their applications in BHI, and the synthesized knowledge of LLMs in BHI (Section 2). Then the methods section outlines our review approaches (Section 3), including data collection and description, topic classification, network analysis, and visualization techniques employed. The result sections are organized in an overall-to-specific manner. First, we provide a two-fold overview (Section 4): the first fold is about content analysis, focusing on research themes and topics; the second one is on network analyses, focusing on scholarly networks and partnerships. Based on the analysis of *research themes and topics*, we further highlight three findings, including (1) *the distributed methodologies* (Section 5), (2) *the diverse prospects of LLM applications* (Section 6), and (3) *specific disease categories* where LLMs have shown promise (Section 7). Finally, the conclusions and discussion section (Section 8) summarizes our key findings, addresses limitations, and provides recommendations for future work in this rapidly evolving field[1].

---

[1] We also provide the workflow and relations among sections in **Appendix I**.

# 2. Backgrounds

The intersection of LLMs and BHI represents a frontier of innovation. To better understand the applications prospects of LLMs in the BHI domain, we conducted a background investigation from three perspectives: (1) *the evolution of LLMs*, (2) *applications of LLMs in the domain of BHI*, and (3) *synthesized knowledge of LLMs in BHI*.

## 2.1 Evolution of Large Language Models (LLMs)

LLMs represent a sophisticated category of language models that utilize neural networks with multi-billion parameter architectures. These models are trained on vast unlabelled textual data using self-supervised learning techniques [10,11]. An earlier milestone was made in 2017 when Google released the Transformer model. This model introduced the self-attention mechanism, which was fundamental for LLMs by capturing contextual relationships and nuanced information among input tokens [12]. Following this model, the introduction of Bidirectional Encoder Representations from Transformers (BERT) in 2018 was another milestone that revolutionized the way that machines understand human language [13].

Later, the evolution of LLMs witnessed a significant moment with the release of OpenAI's GPT-3 in 2020, which has been widely regarded as a game-changer in the field. Having trained using 175 billion parameters, GPT-3's transformer-based model demonstrated an unprecedented capacity for generating text that resembles human writing [14]. This period also gave rise to other significant models such as T5 [15], ERNIE [16], and EleutherAI's GPT-Neo [17], each contributing uniquely to the LLM landscape.

In recent years, the development of LLMs has pivoted towards enhancing both efficiency and contextual understanding. This shift has unlocked more sophisticated and nuanced applications [18,19]. In particular, recent models are not only linguistically adept but also integrate multimodal capabilities, processing both text and other forms of data [20]. This advancement has led to the emergence of various generative AI models, both in closed-source and open-source domains. Prominent closed-source LLMs include ChatGPT by OpenAI [4], Claude 2 by Anthropic [21], and Gemini by Google [22]. Typical models in the open-source domain include LLaMa 2 by Meta [23], and Phi-family models by Microsoft [24].

## 2.2 Applications of LLMs in BHI

In the early stage, NLP applications in BHI primarily focused on extracting and categorizing information from electronic medical records and medical literature. These applications aimed to improve information retrieval [25,26], learn semantic relations of clinical text [27], and train word embeddings [28,29]. These early implementations of NLP have set the stage for the integration of sophisticated models that could handle a broader range of linguistic tasks.

With the advancement of LLMs, the scope of NLP in healthcare has expanded dramatically. In particular, the research of the BERT model in BHI has transitioned from rule-based text processing to more advanced applications [30]. One of its notable applications is text classification, where BERT's contextual analysis significantly enhances the accuracy of categorizing clinical notes, research papers, and patient feedback into relevant medical categories [31–34]. The BERT model has been extensively applied in Name Entity Recognition (NER) and relation extraction within the BHI domain [35–37]. In addition, there has been significant progress in fine-tuning the BERT model for specific applications within BHI.

Noteworthy among these are BioBERT and ClinicalBERT, introduced by [38] and [39], respectively.

Compared to BERT models, the advanced LLMs have shown general-purpose capabilities, which enable them to excel across a broad set of NLP tasks in BHI [40], rather than being designed solely for a single NLP task, such as NER or text classification. For example, LLMs have shown potential in interpreting complex patient data and suggesting medical diagnoses [41–45]. This capability is useful for synthesizing unstructured patient information and supporting clinical decisions. They are also integral to drug-disease identification and drug discovery, where they have shown promise in identifying drug candidates and their effects [46,47]. In addition, the customization abilities of LLMs have unlocked new possibilities in medical education [48–51]. These models could adapt to the learning pace and style of individual students, providing personalized learning experiences.

Among these applications, there are several studies to highlight. For example, [52] evaluated the performance of ChatGPT on the United States Medical Licensing Exam (USMLE). Their findings revealed that ChatGPT achieved scores at or near the passing threshold across all three sections of the exam without any training or reinforcement. [1] proposed an approach for the evaluation of LLMs in the context of medical question answering. Their study showed the promise of LLMs in clinical knowledge and question-answering capabilities.

## 2.3 Synthesized Knowledge of LLMs in BHI

Several review papers of applications of LLMs in BHI have appeared [40,53–57]. We present an overview of the represented review papers in **Table 1**. Two of the earliest review papers of applied research of LLMs in BHI surveyed how LLM applications could be developed and leveraged in clinical settings [40, 57].

As a systematic review of ChatGPT in healthcare, [53] selected papers on Pubmed with keywords 'ChatGPT'. Two-sided taxonomy (application-oriented and user-oriented) were provided to categorize three levels papers (generic comment about the applications in healthcare as Level 1, one or more example uses in specific medical specialty as level 2, and qualitative and quantitative evaluation of ChatGPT in a specialty as level 3). The comprehensive survey [54] particularly focused on the areas of biomedical information retrieval, question answering, medical text summarization, information extraction, and medical education. Their study found significant advances made in the area of text generation but modest advances in other applied research in BHI, such as multi-modal LLM. Moreover, they selected papers with keywords LLM and ChatGPT only within Pubmed. Some emergent research may be omitted because of the limited scope of Pubmed.

Additionally, although the review papers about LLMs involved multiple electronic resource libraries [54,55], applied research of LLMs in healthcare was only one aspect of their broader research. Conversely, [56] solely focused on the applied study of LLMs in investigating Electronic Health Records (EHRs). They categorized 329 papers on OpenAlex with LLM keywords (LLM, Bert, et al. & Electronic Medical Record, et al.) into seven major topics: named entity recognition, information extraction, text similarity, text summarization, text classification, dialogue systems, and diagnosis and prediction. However, concentrating only on EHRs might not fully explore the broader impact and versatility of LLMs in various facets of healthcare, including clinical decision support, medical imaging analysis etc.

Our survey of 1698 papers with bibliometric analysis offers several distinct advantages by providing a more comprehensive and systematic examination of the current state of applied research of LLMs in BHI.  We employ a hybrid approach that not only offers a panoramic overview of the field but also facilitates a detailed exploration of specific research themes. This includes both general LLMs research themes and their applied research on major diagnostic categories within BHI domains. By integrating a bibliometric analysis, we could be able to quantify and visualize trends, research hotspots, and the impact of various studies, providing a data-driven perspective that enhances the depth and rigor of our review.

Another key advantage of our survey is its dedicated focus on emerging LLMs, specifically the ChatGPT model family. This allows us to delve deeply into the unique characteristics and capabilities of these models, which are at the forefront of technological advancement in natural language processing. By concentrating on these state-of-the-art models, we provide valuable insights that are directly relevant to the current and future applied studies of LLMs in BHI. Recently, multimodal LLMs are emergent in the domain of BHI, which integrate and process multiple modal data types such as text and images, offer significant potential for more comprehensive and accurate data analysis, diagnosis, and personalized treatment planning. Our survey highlights the transformative potential of these multi-modal models and underscores the need for further exploration and application in the field of BHI.

**Table 1**: Representative review studies using NLP in the domain of biomedical informatics

| Paper | Type of Review | Paper Count | Scope | Contribution | Limitations | **Bibliometric Analysis** |
|---|---|---|---|---|---|---|
| [40] | Commen tary | None | None | One of the earliest reviews; discusses mixed results of LLMs in medical contexts | No systematic paper collection and analysis framework; subjective opinion potentially introduces bias | No |
| [57] | Commen tary | None | None | Describes opportunities and pitfalls associated with employing LLMs in biomedical research | No systematic paper collection and analysis framework; subjective opinion potentially introduces bias | No |
| [53] | Systemat ic Review | 58 | Paper on Pubmed with keyword "ChatGPT" | Provides a 3-level Application- and user-oriented taxonomy | Limited scope (only research about ChatGPT); could omit most emergent | No |

| | | | | | research by only including publications on Pubmed | |
|---|---|---|---|---|---|---|
| [54] | Comprehensive Survey | 582 | Paper on Pubmed with keywords 'large language models' / 'ChatGPT' | Comprehensive analysis of the diverse applications, including information retrieval, question answering, medical text summarization, information extraction, and medical education. | Limited scope (missing some important applications such as multi-modal LLM); could omit most emergent research by only including publications on Pubmed | No |
| [55] | Bibliometric Review | 5752 | Paper on Web of Science with keywords: (("large" / "big" / "massive") AND ("language model" / "language models")) AND ("BERT" / "GPT-1" / "GPT-2" / "GPT-3" / "ChatGPT") | Identifies patterns in research paradigms, collaboration networks, and thematic trends in LLM research, covering core algorithm developments, NLP tasks, and diverse applications across fields such as medicine, engineering, social sciences, and humanities. | Not dedicated to review papers in the field of BHI; review could be too broad | Yes |
| [56] | Systematic Review | 329 | Paper on OpenAlex with (Large language model / | Categorizes and discusses the reviewed papers into seven major | Limited amount of papers included | No |

| LLM / BERT / RoBERTa / T5 / XLNet / Mistral / Mixtral / Falcon / Qwen / BLOOM / Vicuna / LLaMA / GPT / Claude / Bard / Google PaLM / Gemini) & (Electronic Medical Record / Electronic Health Record) | topics: named entity recognition, information extraction, text similarity, text summarization, text classification, dialogue systems, and diagnosis and prediction |
| --- | --- |

Our investigation into the existing review papers highlights a research gap in the literature: there remains a need for a survey that encapsulates the full spectrum of LLM developments and their specific applications. Our review paper stands out for its multifaceted contributions. Firstly, it offers a detailed survey and bibliometric analysis of the latest LLM applications in BHI, providing a perspective on the evolving trends and challenges within this field. Secondly, the data-driven nature of our review allows for a deeper understanding of the interdisciplinary connections within the published literature and assists in locating the key contributors through semantic network analysis. Thirdly, unlike previous reviews that may have concentrated on particular facets, our work presents a holistic perspective on the trajectory of LLMs in BHI, elucidating how these models have both shaped and been shaped by the needs and advancements in biomedical sciences and health practices. Our contributions are as follows:

1. Our review offers a detailed survey and bibliometric analysis of the latest LLMs' applied research in BHI, providing a perspective on the evolving trends and challenges within the BHI field.
2. The data-driven nature of our review allows for a deeper understanding of the interdisciplinary connections within the published literature and assists in locating the key contributors through semantic network analysis.
3. Unlike previous reviews that may have provide overview of LLM on multiple fields (e.g. engineering, humanities) [55], or one particular domain within BHI (e.g. EHR) [56], our work presents a holistic perspective on the trajectory of LLMs in BHI, elucidating how these models have both shaped and been shaped by the needs and advancements in biomedical sciences and health practices.

# 3. Methods

In this section, we provide an overview of the methodologies employed in the review, which includes data collection and description, topic classification, network analysis and visualization techniques.

## 3.1 Data Collection and Analytics Workflow

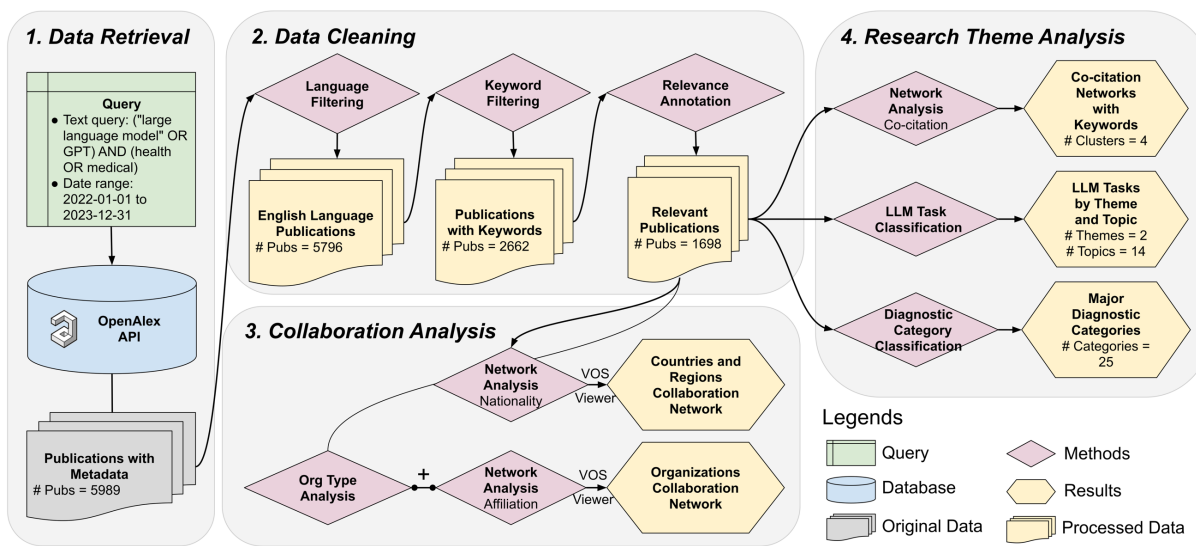

**Figure 1**. Data Selection and Analytic Workflow

**Figure 1** shows the data retrieval, cleaning, and analysis workflow. In this study, the primary data source is OpenAlex, a comprehensive database known for its extensive collection of academic publications. OpenAlex includes both published papers and preprints on platforms like arXiv and medRxiv. This feature allows us to access a broader range of research, including early-stage findings and contributions yet to be peer-reviewed, thereby enriching our dataset with a wider variety of scholarly work. The specific query[2] employed to extract relevant data was:

`("large language model" OR GPT) AND (health OR medical) + [2022-2023]`

This query was chosen to ensure the inclusion of relevant documents that discuss or mention LLMs, including GPT, in the context of health or medical fields. The time frame of 2022-2023 was selected to gather the most recent and relevant insights[3]. Importantly, the decision to avoid explicitly including the model names such as "llama" in the query was deliberate. While "llama" is a term associated with certain models (i.e., Llama by Meta) [23], it also commonly refers to the animal. Including it could dilute the relevance and focus of the research. By structuring the query in this manner, we were able to efficiently isolate documents that are specifically relevant to the intersection of LLMs like GPT and health or medical studies, without the interference of unrelated topics.

---

[2] We conducted the data collection on 01/05/2024.

[3] Some papers that were officially published in 2024 had their original versions published on arXiv in either 2022 or 2023.

Following this, we implemented a more focused local restriction based on the English-language papers and terms "large language model" or "GPT". According to the information on the OpenAlex help page, the API scans through titles, abstracts, and full texts of documents while searching. However, it employs techniques like the removal of stop words and the use of stemming (specifically, the Kstem token filter) to enhance search results. Although these techniques are generally effective, they could sometimes lead to the inclusion of non-relevant documents, particularly after the stemming process. To counteract this issue, we performed a second round of cleaning, aiming to retain only those documents that explicitly mention the model query terms in their titles and abstracts. This step was crucial in refining the results to ensure the relevance and precision of our dataset.

The final step in our filtering process involved the removal of irrelevant papers through human annotation. Even with the advanced algorithmic filtering, some false positives—particularly non-health and non-medical articles—may be retained. To address this issue, we engaged two human annotators who independently reviewed the dataset. Their task was to identify and eliminate any remaining irrelevant papers. After this independent annotation, we measured the agreement rate between the two annotators, which stood at 96%. This human element of the filtering process was vital in ensuring the highest possible accuracy and relevance of the final collection of 1,698 papers for our research.

## 3.2 Topic Classification for Content Analysis

### 3.2.1 RoBERTa Text Classification

For the paper topic classification task, we employed the `roberta-large-mnli` model, a pre-trained transformer-based neural network designed for natural language understanding tasks. While unsupervised methods like topic modeling are generally valuable for exploratory analysis, we've empirically tried a SBERT-based topic modeling method, named BERTopic, but the specificity of the BHI domain made the general semantic-based unsupervised clustering hard to distinguishing topics [58,59]. Another reason for the challenge with SBERT-based models is that the clusters are not as distinguishable for closely related tasks within one field compared to more distinct topics across multiple fields, such as engineering and social science [60]. This lack of clear differentiation further complicates the effective classification of BHI research topics using unsupervised methods. Additionally, SBERT methods typically do not classify a single paper into multiple categories, which can be a significant limitation given that research papers often span multiple topics [61].

Instead, supervised models such as RoBERTa offer enhanced precision for well-defined categories. Specifically, we choose `roberta-large-mnli` for its high performance on the Multi-Genre Natural Language Inference (MNLI) benchmark and capability to leverage pre-trained knowledge, which makes it well-suited for zero-shot learning tasks [62–65]. This model is especially adept at categorizing LLM research papers, which may encompass multiple topics within a single document.

The zero-shot classification process involved defining a set of target topics related to LLMs, such as "model evaluation", "sentiment analysis", "education", and "ethics". There are 14 topics in total, selected by combining research themes of prominent NLP conferences, such as Empirical Methods in Natural Language Processing (EMNLP) and Association for Computational Linguistics (ACL). The final topic list was reviewed by three researchers independently. The purpose of the selection of these topics was to capture a wide spectrum of impactful applied research of LLMs in BHI. We have carefully curated the 14 topics from the original sub-domain lists of EMNLP and ACL. In addition, while these topics may not cover every aspect of the literature corpus, they represent key areas of interest and innovation in the applied research of LLMs in the BHI field.

Using the `roberta-large-mnli` model, each title and abstract was classified into one or more of the 14 predefined topics. The model inferred the relevance of each topic to a given text by predicting the likelihood that the text would be a hypothetical premise for a human-written hypothesis representing each topic[4]. To select the most likely set of predefined topics, we restrict the likelihood to be above 0.1[5].

### 3.2.2 Major Diagnostic Categories

To evaluate the applied research of LLMs in medical domains, we extracted the specific diseases and symptoms from paper abstracts and grouped them into their corresponding Major Diagnostic Categories (MDC). The MDC is a system of classification that organizes diseases and medical conditions into 25 mutually exclusive diagnosis areas that are related to the affected organ system or the etiology of the condition. As the diseases and symptoms mentioned in the abstract directly align with the specific research objectives or questions each study aims to address, this process classifies research papers into their corresponding broader diagnostic categories[6]. For example, epilepsy, Parkinson's disease, and Alzheimer's disease are under "Nervous System" disorders.

Specifically, we employed a multi-step approach to categorize diseases mentioned in abstracts, ensuring accuracy and reliability with collaborative and systematic methods. First, two researchers with biomedical backgrounds reviewed the abstract and identified mentions of disease, disorder, symptoms, and public health crises. Following the identification phase, another pair of researchers group the identified diseases, disorders, and symptoms into their corresponding MDC. Then to ensure the reliability and consistency of the categorization process, an intercoder reliability check is performed with Cohen's Kappa of 0.9. We then include a third annotator, who is an experienced researcher in the BHI fields, to judge the annotation result and resolve discrepancies in data labeling.

## 3.3 Network Analysis Algorithm and Visualization

To construct the bibliometric networks, we employed the VOSviewer [66] software. These networks' entities include organizations, researchers, or individual publications, and the analysis is based on co-citation, bibliographic coupling, or co-authorship relations. VOSviewer utilizes a clustering algorithm based on the Visualization of Similarities (VOS) technique, which effectively maps and visualizes complex bibliometric networks. This algorithm begins by calculating the similarity between items (such as publications, authors, or journals) based on criteria such as co-citation or co-authorship. These similarities then form a matrix, which is used to spatially arrange items that reflect their mutual similarities. Leveraging modularity-based techniques, the algorithm groups items into clusters, which allows for an intuitive representation of the relationships and patterns within BHI. In each network, the size of the node represents total link strength[7], indicating the cumulative strength of the connections an entity has with entities. The edge represents the connections or links between the nodes, illustrating the specific relationships such as co-citation, bibliographic coupling, or co-authorship.

---

[4] In our analysis, the hypothesis is "The topic of this paper is {}." The classification did not require any fine-tuning or training on a labeled dataset, as the model leveraged its pre-trained knowledge to make inferences about the unseen topics.

[5] We tested various thresholds by sampling 100 papers to manually inspect relevance. The threshold of 0.1 was chosen to balance between specificity and sensitivity in the zero-shot classification process.

[6] For detailed disease to MDC mapping, refer to **Table 2**.

# 4. Mapping the Terrain: An Overview of the Diverse Ecosystem of LLM Research in BHI

This section delves into the comprehensive landscape of LLM research within the realm of BHI. Our exploration is structured into two sections: first, the core research themes and topics employing LLMs, and second, the scholarly networks and partnerships that facilitate this research. Through the overview, we identify representative papers that exemplify significant developments and findings. These selected papers are discussed in subsequent results sections (Section 5, 6, 7) to highlight their contributions and innovations.

## 4.1 Research themes and topics

In the burgeoning field of BHI, LLMs have emerged as pivotal tools, enabling the transformation of data into actionable insights. As shown in **Figure 2**, the keyword co-occurrence network adeptly represents the diverse research themes and topics that converge in this multidisciplinary domain[8]. At the center of this complex network lies the interdisciplinary interplay between technologies and BHI fields: Social Science (Cluster 1: blue), Computer Science (Cluster 2: red), Biomedical Science (Cluster 3: green), and Psychological Science (Cluster 4: yellow). Their synergy illustrates the multifaceted nature of the application of LLMs in BHI research.

---

[7] The mathematical definition of total link strength is provided in **Appendix II**.

[8] **Appendix II** shows the top 50 keywords in the network ranked by the total link strength in descending order.

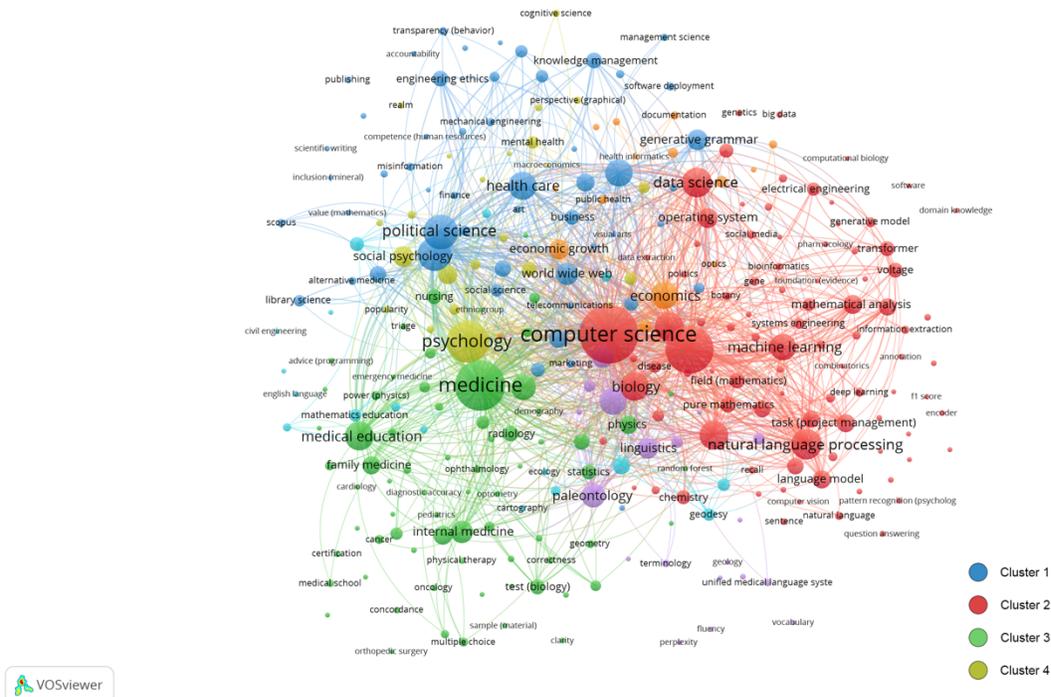

**Figure 2**. Keyword Co-occurrence Network

Cluster 1 highlights the social implications of deploying LLMs in the biomedical and health sciences, including terms such as "engineering ethics," "data transparency," and "knowledge management," which are indicative of a keen awareness of the social dimensions intrinsic to the deployment of technology in sensitive fields. Cluster 2 is strongly associated with the core technical disciplines of LLMs, such as computer science and mathematics. This cluster's prominence underscores a significant research focus on the theoretical and computational foundations which are necessary for the development and refinement of LLM algorithms. The high level of connectivity within this cluster suggests a concerted effort toward advancing the capabilities of LLMs in handling and interpreting complex biomedical data. Cluster 3 emphasizes the potential practical medical applications of LLMs and encompasses various medical specialties and fields such as internal medicine and medical education. This cluster signifies the prospective role of LLMs in clinical practice, medical training, and patient care. Cluster 4 shows concepts at the crossroads of psychological science and its applications within the biomedical and health sectors. This cluster signifies an emerging trend where LLMs have been used to obtain insights into patient psychology, public health, and the societal impact of health interventions.

Overall, this keyword network provides an overview of the state-of-the-arts of LLMs application in BHI. It shows the main topics being studied and the interdisciplinary collaborations that are crucial for making progress in this field. The following sections would examine each of these topics in depth, explaining their contributions and highlighting the interconnected research efforts that could drive the continued advancement of BHI.

### 4.1.1 LLM Research Themes

The categorization of tasks associated with LLMs in the context of BHI into Methodology and Outcome is a strategic way of organizing the research papers' focus areas[9], which delineates between technical development and practical applications/evaluation. **Figure 3** shows the number of papers within each research theme, with red bars indicating the Outcome theme and blue bars indicating methodologies.

In terms of Methodology (blue), LLM topics such as Information Extraction, Inference, Summarization, Sentiment Analysis, and Named Entity Recognition show nuanced capabilities of LLMs in processing and analyzing textual data, which could support various aspects of clinical and research activities in the biomedical sector. The topic of Multilinguality and the topic of Text Generation are also well-represented, illustrating the technical versatility of LLMs and their potential for creating understandable medical content in multiple languages, which is vital for diverse patient communication and international research collaboration. From a technical standpoint, the topic of Image, Vision, Video and Multimodality acknowledges the integration of LLMs with other data forms, which is an important step towards comprehensive analytics in diagnostics and patient care.

For Outcome (red), the highest number of papers centered on the Model Evaluation category, which suggests that there is a significant emphasis on validating and testing the effectiveness and reliability of LLMs within the biomedical field. Model Evaluation is critical because the outputs of such models often inform decision-making in health-related matters where accuracy is paramount. Other LLM tasks in the Outcome category include Dialogue and Interactive System, Education, and Scholarship and Manuscript Writing, representing the substantial interest in using LLMs to distill medical information from various data sources to enhance patient interaction, medical education, and research. The topic of Ethics also has a dedicated focus, which is crucial given the sensitive nature of medical data and the implications of AI in healthcare decisions.

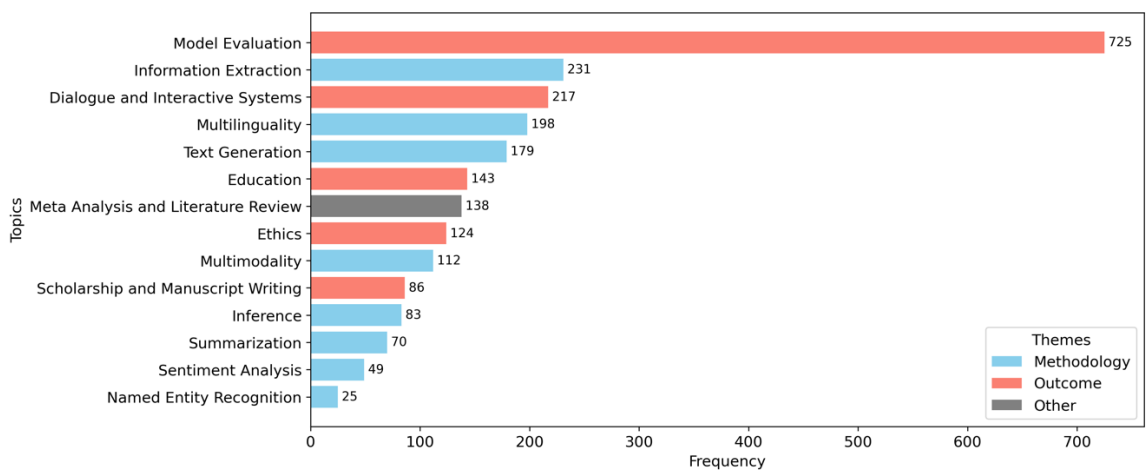

**Figure 3**: LLM Tasks by Research Theme and Topic[10]

---

[9] In **Appendix III**, we present the representative papers for each LLM task.

[10] We include "Meta Analysis and Literature Review" to classify the papers. However, since it is not within the scope of LLM Methodology or Outcome, detailed analysis of papers of this category is presented in **Appendix IV**.

## 4.1.2 Major Diagnostic Categories

**Table 2** categorizes the research papers according to the health issues they address, showcasing the wide-ranging capabilities and applications of LLMs in BHI. Research has predominantly focused on mental health conditions, including depression and ADHD. Similarly, diseases of the nervous system also attract considerable attention, with studies covering disorders from Parkinson's to Alzheimer's disease. The application of LLMs in tracking and managing infectious and parasitic diseases, such as complications from infections and COVID-19, underscores their importance in infectious disease surveillance, particularly in light of recent global health emergencies. Furthermore, research on the circulatory system targets widespread conditions such as heart disease, which continues to be a leading cause of death globally. Other less-represented diseases, such as those affecting the musculoskeletal and endocrine systems, metabolic and digestive disorders, and urinary tract issues, demonstrate LLMs' versatility in tackling a broad spectrum of chronic and acute health challenges.

| **Table 2**: Major Diagnostic Categories Count with Examples | | |
|---|---|---|
| **Category** | **# Collected Papers** | **Examples** |
| Mental Diseases and Disorders | 89 | depression, post-traumatic stress disorder (ptsd), attention-deficit/hyperactivity disorder (adhd) |
| Nervous System | 87 | epilepsy, vestibular schwannoma, carpal tunnel syndrome (cts), parkinson's disease, alzheimer's disease |
| Infectious and Parasitic Diseases and Disorders | 60 | post-infectious complications, signs and symptoms of adverse events following immunization (aefis), symptomatic covid-19 infections |
| Skin, Subcutaneous Tissue, and Breast | 41 | breast cancer, melanoma, skin disease |
| Circulatory System | 40 | congenital heart disease, atrial fibrillation, heart failure |
| Musculoskeletal System and Connective Tissue | 31 | shoulder impingement syndrome, anterior cruciate ligament (acl) injury, rheumatology-related diseases, osteoarthritis (oa), gout |
| Endocrine, Nutritional, and Metabolic System | 28 | anorexia, thyroid cancer, type 2 diabetes mellitus, diabetes |
| Digestive System | 26 | colorectal cancer, inflammatory bowel disease (ibd), inflammatory bowel disease, digestive diseases |
| Eye | 24 | primary acquired nasolacrimal duct obstruction,myopia, cataract |
| Respiratory System | 19 | lung cancer, asthma, metastases, non-resolving pneumonia |
| Hepatobiliary System and Pancreas | 17 | cirrhosis, hepatocellular carcinoma, liver cirrhosis, liver disease |
| Kidney and Urinary Tract | 16 | urolithiasis, end-stage renal disease, transplant chronic dysfunction, graft loss, urinary tract infection (uti) |
| Blood and Blood Forming Organs and Immunological Disorders | 16 | sickle cell anemia, chronic myeloid leukemia, non-hodgkin lymphoma, acute bleeding, anemia severity |
| Ear, Nose, Mouth, and Throat | 13 | oral potentially malignant disorders (opmds), necrotizing otitis externa, neoplastic rhinopharyngeal lesion |
| Female Reproductive System | 12 | infertility, ovarian cancer |
| Alcohol/Drug Use or Induced Mental Disorders | 11 | substance use disorders, drug abuse, addiction, smoking cessation, addiction |
| Male Reproductive System | 10 | prostate cancer, erectile dysfunction |
| Factors Influencing Health Status | 9 | drug-drug interaction (ddi) |
| Pregnancy, Childbirth, and Puerperium | 7 | postpartum hemorrhage (pph) |
| Injuries, Poison, and Toxic Effect of Drugs | 5 | acute organophosphate poisoning |
| Multiple Significant Trauma | 5 | joint contractures, internal organ dysfunction |
| Newborn and Other Neonates (Perinatal Period) | 4 | neonatal diseases |

| Myeloproliferative Diseases and Disorders (Poorly Differentiated Neoplasms) | 4 | chronic myeloproliferative neoplasms |
|---|---|---|
| Burns | 4 | 1st degree burns |
| Human Immunodeficiency Virus (HIV) Infection | 3 | HIV |

## 4.2 Scholarly networks and partnerships

The visualization of the citation network shown in **Figure 4** offers a detailed perspective on the emergent field of LLMs in healthcare. The network includes 312 papers, each with at least five citations, which ensures that the visualization emphasizes the more influential and recognized studies within the field. The structure of the network indicates a close connection among studies, with certain seminal papers emerging as central nodes with their high citations. [50,52,67] are particularly prominent, suggesting their work on model performance evaluation and systematic literature reviews has been widely recognized across the field. Additionally, the network shows the emergence of subfields or specialized areas of research, as illustrated by distinct clusters. For instance, Cluster A (cyan) highlights the focus on radiology reports [68–71], whereas Cluster B (blue) is dedicated to the educational applications within medical specialties, such as dentistry [72–74].

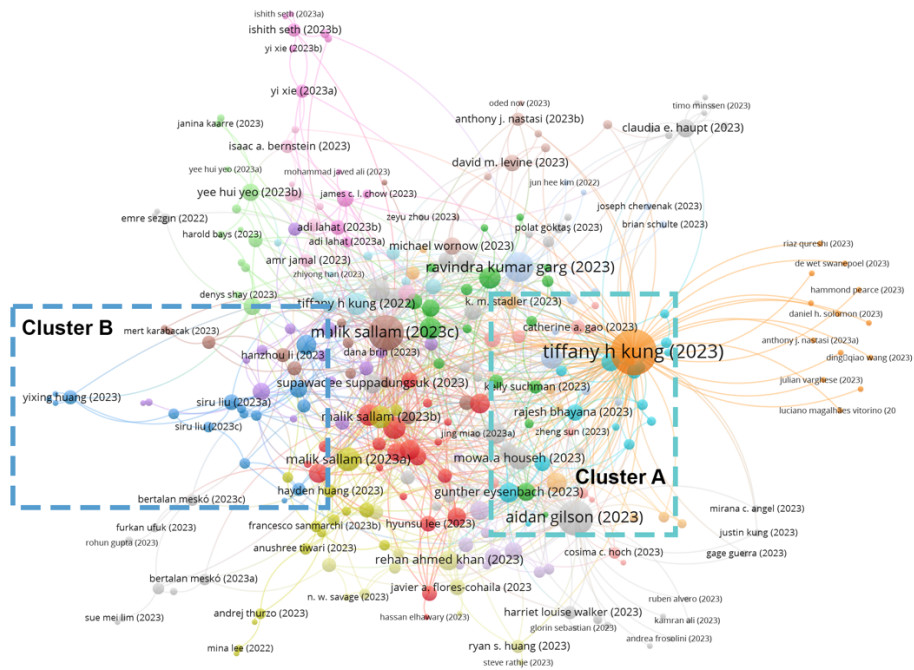

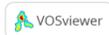

**Figure 4**: Paper Co-Citation Network

The dynamic and collaborative nature of the citation network indicates the ongoing development within this field of study. New theories and methodologies are continuously being integrated. This dynamic is typical for an emerging field, where the foundational work is still being established and where there is significant potential for discoveries and applications.

## 4.2.1 Organization Collaboration Network

The network map in **Figure 5** provides a visual representation of the co-authorship links (with more than 5 co-occurrences) that exist among research organizations across the globe. We observe that the nodes are predominantly universities and research institutions. However, the presence of hospitals and healthcare organizations within this network cannot be overlooked; it signals an integrated research approach where the clinical settings play a crucial role in the translation of academic findings into healthcare advancements. The inclusion of these healthcare entities not only diversifies the nature of the collaborations but also enhances the potential for practical, patient-centered outcomes to emerge from these scholarly partnerships.

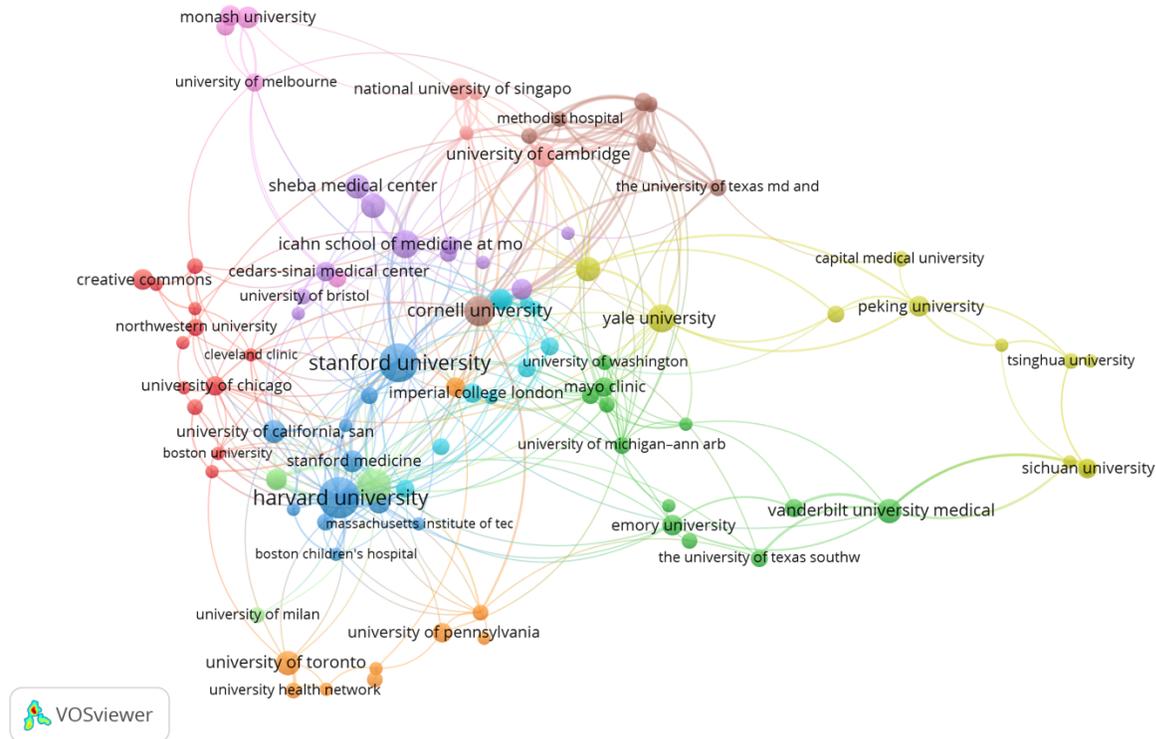

**Figure 5**: Organization Collaboration Network

Certain institutions appear as pivotal nodes within this network. These nodes, often representing universities and research centers like Harvard University, Stanford University, and the University of Oxford, are heavily interconnected with a multitude of other nodes. This finding suggests a high degree of collaborative engagement, which is often a reflection of the institutions' broad research portfolio and its pivotal role in facilitating multidisciplinary studies.

The network also includes tightly interconnected research clusters indicated by colors, suggesting the existence of consortia or research groups that may be working in concert towards a common scientific objective. The network includes edges connecting institutions from multiple continents and countries, which signifies the extent of international collaboration efforts.

## 4.2.2 Collaboration Network among Countries

**Figure 6** provides a visual representation of a collaboration network among countries and regions, with an overlay that indicates the average publication year of papers from each country and region. This visualization not only shows the collaborations that exist among countries but also provides a temporal dimension of how the research landscape has evolved. There are three main findings regarding the early pioneers, the major collaborators, and the dynamic and evolving network.

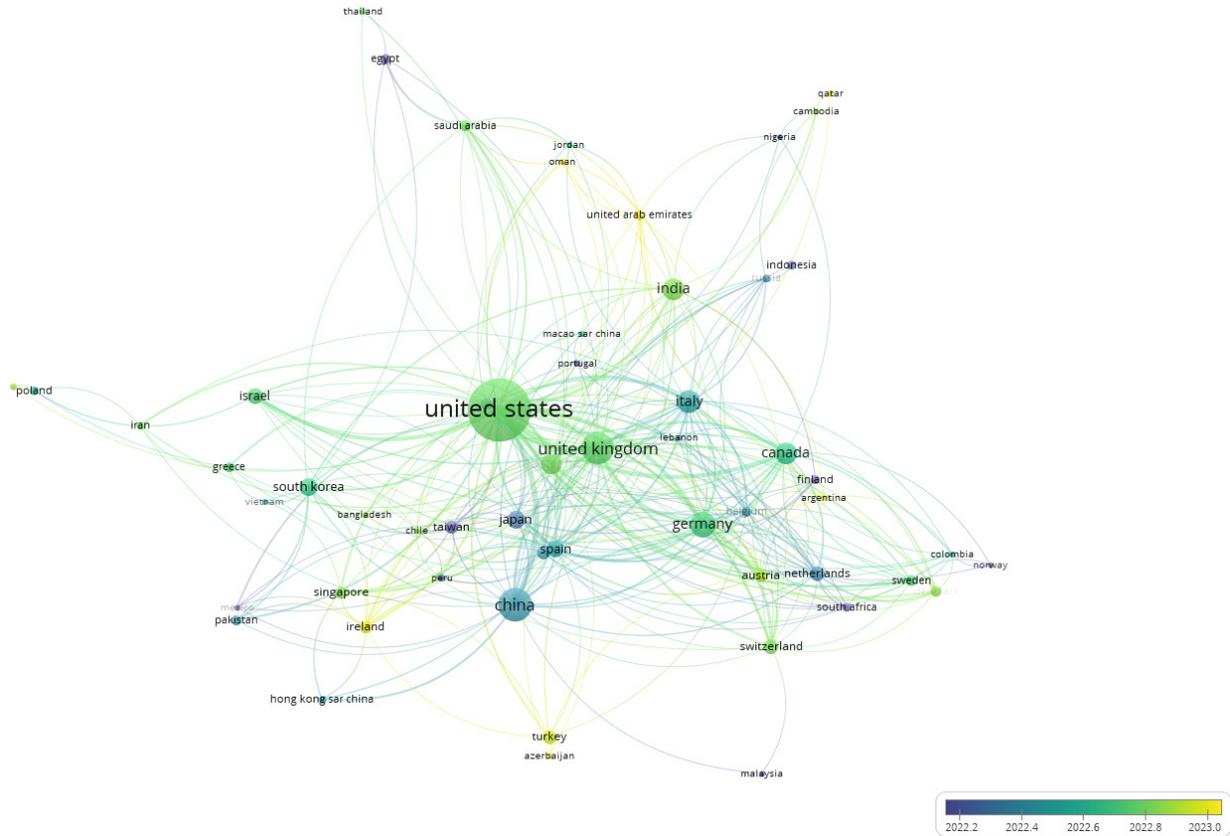

**Figure 6**: Collaboration Network among Countries

**Early Pioneers.** It is shown that countries such as Japan and the Netherlands have begun research on LLMs earlier, making them pioneers in this field. Their early start suggests that these countries have established a strong foundation in LLM research, contributing significantly to the early development and understanding of these technologies in the field of BHI

**Major Collaborators.** As shown in **Figure 6**, the United States and the United Kingdom are depicted with a large total link strength (as indicated by the size of the nodes), which is indicative of their strong influence and the density of their collaborative networks. A large link strength suggests these countries are central nodes in the network, engaging in numerous collaborative research projects, and often being the driving force in pushing the frontiers of LLMs. Their central role in the network underscores their importance in both producing and disseminating LLM knowledge.

**Dynamic and Evolving Network.** The network is dynamic and evolving, with countries like Ireland, Turkey, and the United Arab Emirates emerging as participants. It indicates that the field of LLMs is growing, attracting a diverse set of contributors and expanding the geographic diversity of research. The participation of these countries may bring new perspectives and innovations to the field, and their increasing involvement highlights the global interest and importance of LLM research.

# 5. Navigating the Spectrum: the Distributed LLM Methodologies in BHI

The study of LLMs in the area of BHI covers a wide range of methods and use cases, showing a major change in how AI is used in the biomedical and health fields. This detailed review starts with how LLMs change information extraction, making it easier to handle and understand different types of data, like clinical notes and radiology reports. The discussion then moves on to multilinguality, looking at how well LLMs perform in different languages and the challenges and solutions in creating content in multiple languages. The next parts focus on text generation, highlighting how LLMs play a role in both medical writing and communication with patients. The study also looks at how LLMs can handle multiple types of data, like images and genomic data, which helps improve diagnosis and prediction. As the discussion continues, it emphasizes how LLMs are important for making conclusions and analyzing sentiment, showing their significant impact on understanding complex medical data and human feelings. The study ends with a look at how LLMs are used in named entity recognition, pointing out current progress and potential for future improvements. Overall, each part highlights the diverse and specific applications of LLM methods in changing BHI research and practice.

## 5.1 Information Extraction (including Sentiment Analysis and Named Entity Recognition)

The utilization of LLMs has rapidly reshaped the BHI research landscape, notably in the domain of information extraction. Recent literature underscores their transformative impact across multiple applications, which we will discuss in three main areas: structured information extraction, sentiment analysis, and named entity recognition (NER). Sentiment analysis and NER are sub-tasks of information extraction that play crucial roles in understanding and organizing unstructured data.

For structured information extraction, some studies demonstrate LLMs' proficiency in enhancing diagnostic accuracy in hematology [75], extracting structured information (e.g., diseases, symptoms, and signs) from vast textual data (e.g., clinical notes, EMR notes, and radiological reports) in various languages [76–80], and identifying narrative entities in the news domain [81–83]. In addition, a part of the research illustrated the ability of LLMs in assisting in the extraction of evidence-based explanations and enabling the accurate retrieval of information from clinical documentation, providing support for medical practitioners' decision-making [84–86], [87,88]. Our review also indicated that LLMs are instrumental in extracting medication mentions [89], classifying events and contexts in clinical notes [90], and improving the understanding of medication adherence through the detection of drug discontinuation events from social media data [91]. They also excel in generating structured outputs on medications and temporal relations, further aiding in disease prediction and clinical decision support [92–94]. These advancements, coupled with self-verification techniques [77] and the extraction of demographics and social determinants of health from Electronic Health Records (EHR) [95–97], [98,99] illustrate LLMs' capacity to integrate and analyze healthcare data effectively.

In BHI domains, LLMs contribute to the refinement of sentiment analysis tools [161–163]. For instance, a model utilizing weights from a publicly available zero-shot classifier, which is built from the BART LLM and fine-tuned on the MNLI dataset, has been employed to evaluate linguistic nuances during psychological therapy sessions [164]. Similarly, other research find that LLMs could be used to analyze patient feedback, clinical notes, and public health discussions, thereby gauging public sentiment on health-related matters [165], understanding patient experiences [166], monitoring mental health trends [167,168], and identifying cognitive distortions or suicidal tendencies [169]. Additionally, LLMs in sentiment analysis facilitate medical education [163,170–172] by fostering interactions between medical trainees and educators, detecting thematic differences and potential biases, and revealing how feedback language may reflect varying attitudes towards learning and improvement [173]. LLMs could also contribute to sentiment analysis of research articles and medical journals, offering insights into the research community's responses to novel findings or treatments [174,175].

Moreover, LLMs have been applied to improve the efficiency and performance of Named Entity Recognition (NER) in BHI domains. For instance, LLMs have helped identify ancient Chinese medical prescriptions from the Song Dynasty [100,176]. While there are not too much representative literature compared to other methodology subdomains, [177] identifies the needs of further developing supervised medical NER models, especially when human-annotated data that are unavailable.

## 5.2 Multilinguality

Our review highlights the emerging research applications of multilingual LLMs in BHI. Some research has explored how to use multilingual LLM to generate multilingual contents in BHI. The content generation tasks include using multilingual LLMs for dataset generation [100–102] and question generation [100–102][11]. Multilingual LLMs are also leveraged to identify personal health information in Chinese-English code-mixed clinical text and ancient Chinese medical prescriptions [103,104]. These studies demonstrate the versatility and potential of multilingual LLMs in processing low-resource multilingual and cross-cultural biomedical and health information.

Other research papers concentrate on evaluating LLM performance across various languages, including English, Korean, Spanish, Turkish, and Chinese [105–111]. Studies explore multilingual question answering, using the Japanese National Examination for Pharmacists (JNEP) [112], Korean dermatology specialty certificate examination , and Persian medical residency examination [108]. LLMs, including ChatGPT, were also tested for their ability to generate multilingual health-related questions [100], and ability to facilitate multilingual communication [113]. By comparing the results obtained from different language settings, these studies focus on the correctness, consistency, and verifiability of LLMs' responses.

## 5.3 Text Generation

Research LLMs-based text generation in BHI concentrates on two main purposes: medical scientific writing and clinical patient-facing writing.

In medical scientific writing, current research on text generation predominantly focuses on two areas. The first area focuses on the potential utility of LLMs, particularly GPT-4, as tools for authoring various scientific publications. The general consensus is that human-written texts are more concrete, diverse, and typically contain more useful information [114–116]. In contrast, medical texts generated by GPT-4 prioritize fluency and logic, often using general terminologies instead of context-specific information [114,117]. AI-generated texts may include inaccurate information, fabricated references, and lack the inclusion of recent literature [118–120].

---

[11] The generation tasks here exclude text generation, which is discussed in Section 5.3.

The second area is the effectiveness of distinguishing LLM-generated texts through human evaluation or AI-driven Output Detection mechanisms. Some studies focus on detecting AI-written text in specific sections of BHI papers, such as the abstract and background [121,122]. While LLM-based methods are generally useful in distinguishing AI-written abstracts from original ones, they struggle in the field of radiology where both human reviewers and Output Detectors fail to differentiate GPT-generated abstracts from original ones [122]. It has also been claimed that distinguishing AI-written backgrounds from human-written ones is challenging [122]. More robust Output Detectors have been developed to distinguish AI-generated text from human-generated text [123,124]. Overall, researchers advocate for chatbots to serve as assistants rather than authors in scholarly work, emphasizing the importance of transparency if chatbots are involved in generating academic content [125].

For clinical patient-facing writing, efforts have been made to evaluate the feasibility of using GPT-4 for generating case reports and responses to various patient inquiries about surgical procedures and health-related matters. These include responding to postoperative questions [126], generating health messages [127], aesthetic surgery advice [128], pro-vaccination message generation [129], and communication in palliative care [8]. Most studies show positive results regarding GPT-4's ability to generate coherent, easily comprehensible answers. One study even suggests that AI-generated messages are comparable to human-generated ones in terms of sentiment, reading ease, semantic content, and suggestions [127]. However, its accuracy, completeness, and extent of personalization still need improvement [128]. Therefore, AI models cannot replace a human agent at present [8].

## 5.4 Multimodality

Multimodality in large language models within BHI refers to the ability of these models to understand and process multiple types of data beyond text, which includes imaging, audio, and genomic data. In our scoping review, papers on multimodal LLMs have been applied to various aspects of BHI, including healthcare in general [130,131], medical image analysis [132–135], radiology [136,137], pharmaceutical sciences [138], dentistry [73], public health informatics [139]. Methods used in these papers can be crudely classified into pretrain-from-scratch [140–146] as well as finetuning based on the pre-trained or instruction-tuned models [147–153] such as Vicuna, SAM, BLIP, Llama, OpenLlama, etc.

As healthcare and medicine is a highly specialized field, many multimodal models are uniquely adapted to enhance tasks in vision, audio, and genomic analysis. In vision applications, models are designed for tasks including image-to-text medical report generation [147,154], [150,155], [156], medical image captioning [142,144,157], medical video retrieval [158], and video anomaly detection [152]. In [156], LLMs integrate Vision Transformers (ViT) and Faster R-CNN with GPT-2 to analyze brain images for dementia, enhancing diagnostic accuracy by capturing intricate visual features and generating detailed textual reports. Specified models are also developed in audio and genomic applications: LLMs such as the Diagnosis of Thought (DoT) model [159] assist in psychotherapy by detecting cognitive distortions from patient speech, and aiding therapists in understanding and addressing mental health issues more effectively. In the fields of genomics, protein language models predict the impact of genetic variations on protein structure and function, identifying potential compensatory mutations in pathogenic variants [160].

## 5.5 Inferences

In addition to LLMs' application in correlational or empirical studies in BHI, they have also been instrumental in inferences, with focuses on analyzing associations and causal relationships. For example, LLMs facilitated a Socratic dialogue with ChatGPT to analyze the causal effects of PM2.5 on human mortality risks. After extensive fine-tuning and addressing confounding factors, a causal link was established [161]. Moreover, LLMs have been adapted to develop a Natural Language Inference system specifically for Clinical Trial Reports. This system focuses on extracting and interpreting medical evidence to enhance the accuracy and reliability of these reports [162]. In a different application, the GPT model has been utilized for medical image analysis. Demonstrating its capabilities as a plug-and-play transductive inference tool, GPT has proven effective in detecting prediction errors and improving accuracy in medical images, highlighting its potential for broader applications in this field [163].

# 6. Expanding the Horizon: the Diverse Outcomes of LLMs in BHI Applications

The integration of LLMs has also expanded the horizons of BHI, leading to a diverse array of outcomes and applications. Beyond enhancing NLP capabilities, LLMs have facilitated a more personalized and nuanced approach to patient engagement, enabling healthcare providers to tailor their communication and interventions based on individual patient profiles through dialogue and interactive systems. In addition, LLMs have revolutionized scholarship and manuscript writing, which are also applicable to BHI fields. Furthermore, the evaluation and ethics assessment of LLMs have become essential research topics in BHI, given the high standards of precision and stability in healthcare and medical systems. This section explores the multifaceted impact of LLMs across various BHI applications, highlighting their potential to revolutionize patient care and medical research.

## 6.1 Dialogue and Interactive Systems

The LLMs have been implemented in the newly developed chat-box as an AI-assistant for a healthcare conversion including personalized health diagnosis and intervention in BHI. Typically, the chat assistant, based on either naïve conversational AI or generative AI systems, was designed to help in the analysis of the message from dialogs [181–184], the estimation and the evaluation of the health status [181], and the generation of the high-quality responses [181,182] after considering the possible knowledge, including the patient's EHRs and medical knowledge in the clinical setting. For example, an LLM-derived chatbot called CareCall [181] was designed to support people and alleviate feelings of loneliness. It leads to frequent open-ended conversations, generates replies by using a pre-trained LLM model, captures the health metrics and emergency alerts, and displays the reports for social works. Another newly developed application powered by the ChatGPT-3.5 model [182] allows advising the callers with up-to-date personalized medical suggestions based on the conversation. In addition, a prospective of using ChatGPT within healthcare, especially during the pandemic period, was proposed which helps with answering the patients' health-related questions [185]. The high-quality performance of using the AI assistant confirms that the models can understand and reply to people's needs. However, privacy, ethics, and information accuracy are the major concerns while the LLM/AIs are involved in generating professional responses regarding disease diagnoses and drug suggestions [185,186]. More rigorous tests are needed to guarantee the safety of using the LLM in clinics [186].

## 6.2 Scholarship and Manuscript Writing

As more researchers in the BHI domain use ChatGPT and other AI-technologies in writing manuscripts, the discussions around the use of LLMs in scientific writing have been emphasized, accompanied by a rise in various concerns. Although LLMs can improve writing quality, summarize relevant articles, and facilitate manuscript translation [187], it faces challenges in accurate referencing [188], unintentional plagiarism, and data biases [189]. Establishing regulations and guidelines for the use of LLMs in scientific writing is crucial for assessing both effectiveness and ethical considerations [48,190].

## 6.3 Education

Researchers have assessed LLMs' abilities to enhance medical education, discussing their potential to improve the current education and decision-making process. LLMs exhibit similar performance in comparison to human achievement without specialized training on both the United States Medical Licensing Exam (USMLE) [52] and more specialized domains such as neurology board-style examinations [191]. LLMs can also enhance student engagement and learning experiences [49], especially personalized curriculum development and study plans [192], albeit with considerations of ethical challenges [49,50], algorithmic bias, and plagiarism [50,192]. Additional efforts are required from educators, students, and model developers to establish clear guidelines and rules for their applications ethically and safely in academic activities [192]. These perspectives on using the LLMs highlight both the potential benefits and ethical considerations surrounding the integration of LLMs in medical education and practice.

## 6.4 Model Evaluation

As demonstrated in the previous sections, LLMs are widely utilized in a range of applications within BHI. To assess their effectiveness, new frameworks, benchmarks, and metrics for evaluating the performance of these models have been developed. Frameworks such as the TEHAI (Translational Evaluation of Healthcare AI) have been proposed by research teams to evaluate the capability, utility, and adoption of such systems in healthcare [193]. Papers also set benchmarks by assessing the performance of LLMs on various tasks [172,194,195], using relevant datasets such as MIMIC for general medical information and OpenI for radiographs [196]. In their evaluation, metrics such as ROUGE-L have been frequently used [197]. In some cases, additional human evaluations are introduced, which rely on qualitative coding of LLM outputs. For example, for LLMs applied to summarize medical evidence [198], human efforts of evaluating the model-generated summaries involve open coding of qualitative descriptions of error types for medical evidence summarization, drawn from qualitative methods in grounded theory. As another example, human evaluation involves recruiting human subjects to interact with chatbots and solicit their responses [199,200].

In **Appendix V**, we present a thorough analysis of the specialized and contextualized model evaluation in specific disease categories. Taking mental health disease as an example, we highlight evaluation techniques in mental health applications against various metrics and datasets.

## 6.5 Ethics

Ethical discussions on LLMs caution against the application of LLMs in high-state contexts and center around issues of misinformation, bias and inequalities, privacy, transparency. [201], [202]. The use of LLMs as a clinical decision support tool as well as service-providing tool through chatbots can potentially harm patients when they make false recommendations, diagnoses, or prescriptions [201,203]. Such harms, while unintended, are rooted in the corpus of training data embedded in unequal social processes [204]. Moreover, those negative consequences can also be compounded when human health professionals' judgments and decision-making processes are influenced by such biased diagnoses [201]. In particular, the use of AI-generated texts or conversational chatbots in medical contexts often involves patient-specific medical information [201,205,206]. This might introduce additional privacy harms to patients, since these technologies often require access to patents' sensitive information and medical record data [207]. For the responsible use of such technologies, clinicians will need to critically review and validate generated texts or outputs before deploying them in practical settings. Besides, the lack of consent sharing poses another concern around data privacy and security in healthcare [208].

# 7. Applying LLMs in Specific Disease Categories: Popular Fields and Open Opportunities

This section provides a detailed exploration of the transformative impact of LLMs on various disease categories, focusing particularly on mental health, nervous system disorders, and other open opportunities. Mental health and nervous system disorders are the top two widely represented topics in the collected corpus, as indicated by the counts in Table 1. We focus on these two areas as examples to analyze the trending LLM-based BHI applications, while uncovering additional domains as open opportunities. By understanding how LLMs can be effectively applied in these well-represented domains, we can extend these insights to other disease categories, thereby broadening the scope and impact of LLM technology in healthcare.

## 7.1 Mental Health

As shown in **Figure 7**, LLMs are poised to revolutionize mental health care by enhancing diagnostic processes, intervention strategies, and overall mental health and well-being promotion. The potential for LLMs in these domains is vast, ranging from facilitating early detection of mental health issues to providing scalable interventions.

**Diagnosis.** The application of AI in mental health diagnostics has been rapidly advanced with tools like GPTFX [209], which exhibits a remarkable ability to classify mental health disorders and generate relevant explanations. This approach not only enhances the performance of mental health disorder detection but also provides valuable interpretability for the predictions, which is a crucial aspect for clinical applications. The study "Advancing Mental Health Diagnostics: GPT-Based Method for Depression Detection" [7] leverages transformer networks like BERT, GPT-3.5, and GPT-4 to analyze clinical interviews. They have shown strong abilities to understand complex linguistic patterns and contextual cues.

These pioneering studies indicate that LLMs could be instrumental in mental health care by providing nuanced, scalable, and efficient tools for diagnosis. By analyzing language with the unprecedented depth and breadth, LLMs could uncover mental health patterns that may be imperceptible to humans, assist in early detection, and offer continuous support for individuals struggling with mental health issues.

**Intervention**. The field of mental health intervention has benefited through the integration of LLMs and digital health technologies. In [170], researchers proposed a mobile app that utilizes GPT technology for tracking psychological mood changes and providing e-therapy. By offering a platform for users to record and analyze their psychological fluctuations, it aids in identifying triggers for negative mood changes, effectively functioning like a virtual therapist. The app's evaluation underscores its efficacy in journaling and basic AI-driven mental health guidance, exemplifying the potential of LLMs in personal mental health management.

Community-based mental health support can also leverage the advanced capabilities of AI and LLMs, providing more healthcare resources. The paper titled "Enhancing Psychological Counseling with a LLM: A Multifaceted Decision-Support System for Non-Professionals" [210] highlights the need for psychological interventions in the social media sphere, where expressions of negative emotions, including suicidal intentions, are alarmingly prevalent. The model leverages the advanced capabilities of AI and LLMs to empower non-professionals or volunteers to provide psychological support. By analyzing online user discourses, the system assists non-professionals in understanding and responding to mental health issues with a degree of accuracy and strategy akin to professional counselors.

These pioneering applications of LLMs in mental health interventions demonstrate their immense potential in both personal and community settings. Supporting nuanced, user-friendly, and scalable solutions, LLMs have reshaped the landscape of mental health care. They offer innovative tools for real-time emotional tracking, mood analysis, and intervention, facilitating broader access to mental health support and enabling effective responses to complex emotional expressions.

**Promotion.** Healthcare promotion, particularly in the realm of mental health and well-being, has undergone a significant transformation with the advent of AI-based Conversational Agents (CAs) [211,212]. The integration of these advanced technologies has not only reshaped therapeutic approaches but also expanded access to mental health resources. This shift is well-articulated in the comprehensive paper titled "Systematic review and meta-analysis of AI-based conversational agents for promoting mental health and well-being" [213]. The study underscores that the quality of human-AI therapeutic relationships, content engagement, and effective communication significantly shape the user experience. It implies that while AI-based CAs could be highly effective, their impact is greatly influenced by the quality of interaction and the relevance of the content they provide.

Additionally, LLMs play a crucial role in healthcare promotion not only by raising overall awareness but also by offering patient-centric recommendations [214]. They effectively address and dispel common misconceptions and myths about mental health, significantly contributing to the reduction of stigma associated with mental health issues. By educating the public in a non-judgmental and informative manner, LLMs help cultivate a more understanding and supportive community. Furthermore, these models are adept at disseminating a wealth of health-related information in formats that are easily comprehensible. They offer insights on a wide range of topics, from general wellness and stress management to the critical importance of mental health. This comprehensive approach aids in heightening awareness and educating people about the importance of maintaining good mental health, as well as recognizing the early signs of potential issues.

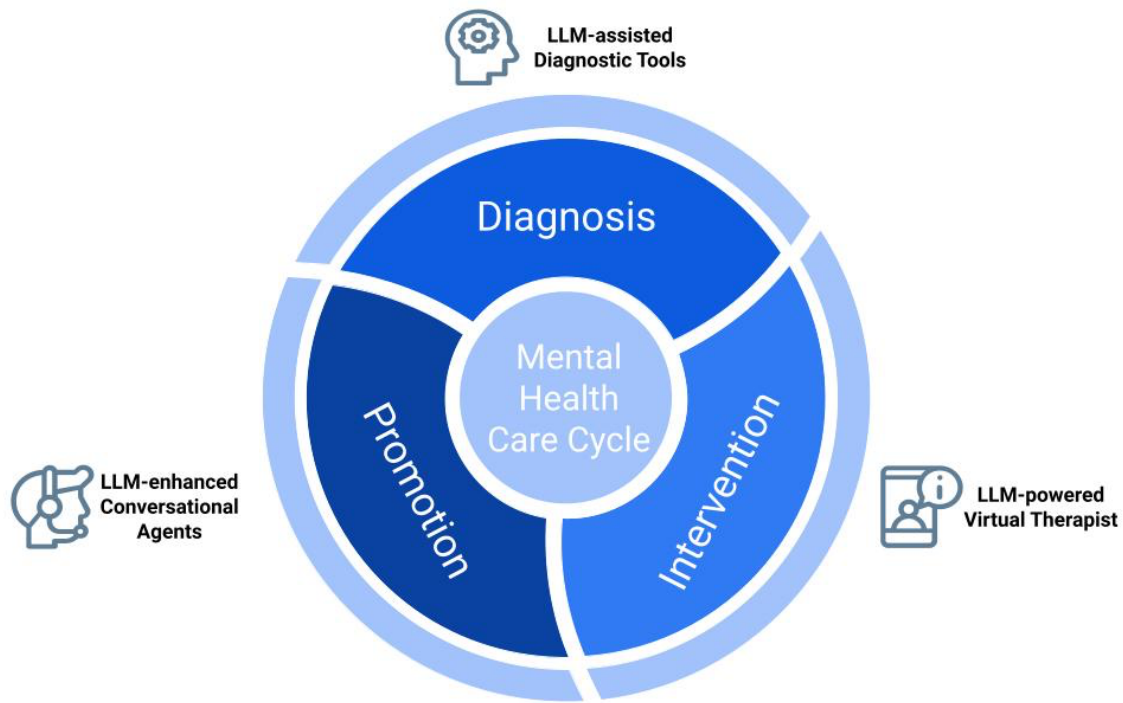

**Figure 7**: Integration of LLM Technology in the Mental Health Care Cycle

## 7.2 Nervous System

In the realm of neurological disorders, leveraging LLMs for disease prediction signifies a groundbreaking shift toward harnessing the intricacies of human language and clinical data. Two pivotal studies exemplify this innovative approach, particularly focusing on multimodal data to predict diseases of the nervous system.

The study, "Predicting dementia from spontaneous speech using large language models," [215] delves into the predictive potential of LLMs by analyzing physicians' clinical notes for signs indicative of seizure recurrence in children following an initial unprovoked seizure. Their work demonstrates that the nuanced understanding captured from Electronic Medical Records could significantly augment the predictive accuracy of seizure recurrence. Another paper "Multimodal Approaches for Alzheimer's Detection Using Patients' Speech and Transcript" [216] ventures into the domain of Alzheimer's Disease detection by employing a multimodal strategy that integrates patients' speech and transcript data. This study underscores the immense potential of multimodal data in enhancing Alzheimer's detection, and sheds light on the complexities and opportunities inherent in leveraging speech data for the prediction of neurological diseases, paving the way for more effective and nuanced diagnostic tools.

Both of the above studies underscore the significant advancements made in the domain of neuroscience, particularly through the use of LLMs and multimodal data analysis. By capturing and integrating diverse data types, from clinical notes to speech and transcripts, researchers could unveil previously obscure patterns and indicators of disease, offering promising new avenues for early detection and treatment strategies for conditions affecting the nervous system.

## 7.3 Open Opportunities

The application of LLMs in BHI holds promising potential to revolutionize disease diagnosis, prediction, and intervention, other than the mental health and neurological disorders that have been extensively researched. Though their use in clinical fields is still in the beginning stages, there exist several opportunities for LLMs to significantly enhance patient care and disease prognosis, particularly in areas such as hospital management, adverse drug reaction, infectious diseases, and health promotion

In clinical settings, LLMs could be instrumental in identifying correlations or even casual relationships [217] by referencing vast datasets such as clinical notes, emergency care reports, and poison control center data. It could lead to the development of more effective triage systems in emergency departments [218] and quicker, more accurate diagnoses [219] . Ultimately, it would help reduce the time needed to administer antidotes or interventions that alleviate symptoms and monitor drug/treatment reactions. Additionally, through the in-depth analysis of the language and semantic information embedded in these full EHRs, LLMs could predict potential personalized treatments [220] to mitigate adverse drug reactions [221].

In the management of infectious diseases with or without pandemic potential, such as sexually transmitted disease (STD), influenza, and COVID-19, LLMs could play a pivotal role in improving patient engagement, promoting adherence to antiretroviral/antibacterial therapy, and monitoring disease progression [222]. By analyzing patient interactions, social media, and support group communications, LLMs could identify language indicative of treatment fatigue or social determinants affecting adherence [98]. Furthermore, through the analysis of clinical narratives over time, LLMs could detect subtle changes in patient status, predict potential comorbidities, and personalize patient education and intervention programs [223]. It could lead to improved health outcomes and quality of life for individuals affected by diseases that currently have no cure.

Finally, LLMs could also extend their contributions beyond disease settings. For example, LLMs can also support the training of medical professionals through simulations and interactive learning platforms, providing personalized learning experiences and improving the quality of medical education [192]. LLM can also benefit public health promotion by enabling more precise and targeted health communication strategies [224].

The potential of LLMs in these medical domains is vast, offering opportunities for enhanced diagnostic accuracy, personalized treatment, and patient care. As the technology and methodologies behind LLMs continue to advance, their integration into clinical workflows and research initiatives would likely become increasingly prevalent, driving forward the capabilities of modern medicine. As the capabilities and applications of LLMs in healthcare expand, there will be a growing need for research into their ethical, legal, and social implications to ensure they are used responsibly and equitably.

# 8. Conclusions and Discussion

Our review has shown important trends and developments in using Large Language Models (LLMs) for Biomedical and Health Informatics (BHI). Applying LLMs has changed the methods and potential outcomes in the healthcare field. Particularly from January 2022 to December 2023, there has been a big increase in the number of research articles, showing rapid progress in this field. These applied research include better diagnostic tools, improved patient engagement, more efficient management of Electronic Health Records (EHRs), and the emerging field of personalized medicine.

The use of LLMs in BHI has captured advanced natural language processing capabilities, greatly improving medical diagnosis, patient care, and research methods. Our network analysis shows that LLMs have also fostered collaborative networks across different disciplines, including academia, healthcare, and technology industries. This multidisciplinary approach is vital for the responsible growth and ethical application of LLMs. Our review also highlights an increasing focus on addressing practical challenges and ethical implications, such as data privacy and AI bias, underlining the need for robust policy frameworks. The potential impact of LLMs in BHI is significant, but it requires a balanced approach considering both the technological capabilities and the ethical, legal, and social implications.

In summary, our review provides a comprehensive resource for stakeholders in the healthcare sector. It offers an overview of the current state of LLMs in BHI and insights into future directions. As LLMs continue to evolve and integrate further into healthcare, understanding their development could be crucial for researchers, clinicians, policymakers, industry leaders, and all stakeholders. It is also important to remain committed to the ethical and responsible use of LLMs in advancing healthcare.

## 8.1 Limitations

This review is subject to certain limitations. First, our classification methodology, while able to conduct multi-label classification, primarily focuses on identifying the most relevant topics within each article. This approach is effective in streamlining the analysis but may overlook the multi-faceted nature of some research papers where secondary topics could also hold significant relevance.

Second, the scope of our review is centered on LLMs, potentially excluding foundation models operated in other modalities such as vision and voice. Additionally, the specific use of biomedical and health-related keywords in our search criteria may have inadvertently excluded relevant studies that do not explicitly use these terms but are pertinent to the field.

Another potential limitation stems from the data-collection process. At the time of our data collection, OpenAlex did not facilitate a refined search based on keyword matches within titles or abstracts. Therefore, we applied several predefined rules, such as filtering articles based on key search terms in the abstracts. We also note a significant portion of the collected papers are preprints, which have not undergone the peer-review process and whose findings and assertions are not established. Although studies, such as [225], have found that over 75% of preprints are eventually published in peer-reviewed journals, we recognize the need for additional validations to ensure the reliability and accuracy of the information presented in these preprints.

These limitations present several opportunities for future work to refine the review. One future work could investigate the application of foundation models in other modalities in BHI fields, including vision and voice. Another future work could continue to collect articles and track the trends in this area.

## 8.2 Future Work

Looking ahead, LLMs have recognizable potential to transform healthcare delivery and patient outcomes. As LLM capabilities continue to evolve, our future work will focus on exploring more advanced ways to integrate LLMs into BHI. This will involve addressing emerging ethical and operational challenges, such as ensuring responsible and fair use of LLMs in healthcare, which is crucial for fully realizing their potential.

The field is evolving rapidly, so ongoing monitoring and analysis will be necessary. We anticipate a surge in publications and citations related to LLMs in the near future. Therefore, continuously updating our review will be essential to maintain its relevance and impact. Our future work will also explore foundation models beyond LLMs, acknowledging the growing importance of multi-modal systems in healthcare. By expanding our research focus, we aim to provide a more comprehensive understanding of the role of advanced computational models in BHI, thereby contributing to the development of more effective and ethical healthcare solutions.

# Declarations

## Authors' Contributions

H.Y. and L.F. contributed to ideal conceptualization and formal analysis; H.Y., L.F., L.L., J.Z., Z.M., L.X., W.H., S.H., M.J. contributed to manuscript writing; H.Y was responsible for producing all figures and tables; Y.Z., A.G. and X.M. are responsible for result validation and draft editing. All authors reviewed the submitted manuscript.

## Funding

This study is not currently funded.

## Availability of Data and Materials

The dataset of the papers analyzed for this manuscript is available from the corresponding author on request.

## Ethics Approval and Consent to Participate

Not applicable.

## Consent for Publication

Not applicable.

## Competing Interests

The authors declare that they have no competing interests.

## Acknowledgments

Not applicable.

(BIBM). IEEE; 2015. pp. 625–628.

Neurology Board-Style Examination. JAMA Netw Open. 2023;6: e2346721.

# Appendices

## **Appendix I**. Paper Organization Workflow

In Appendix I, we provide a visual representation of the workflow used for organizing this research paper.

**Figure 8**. Paper Overview

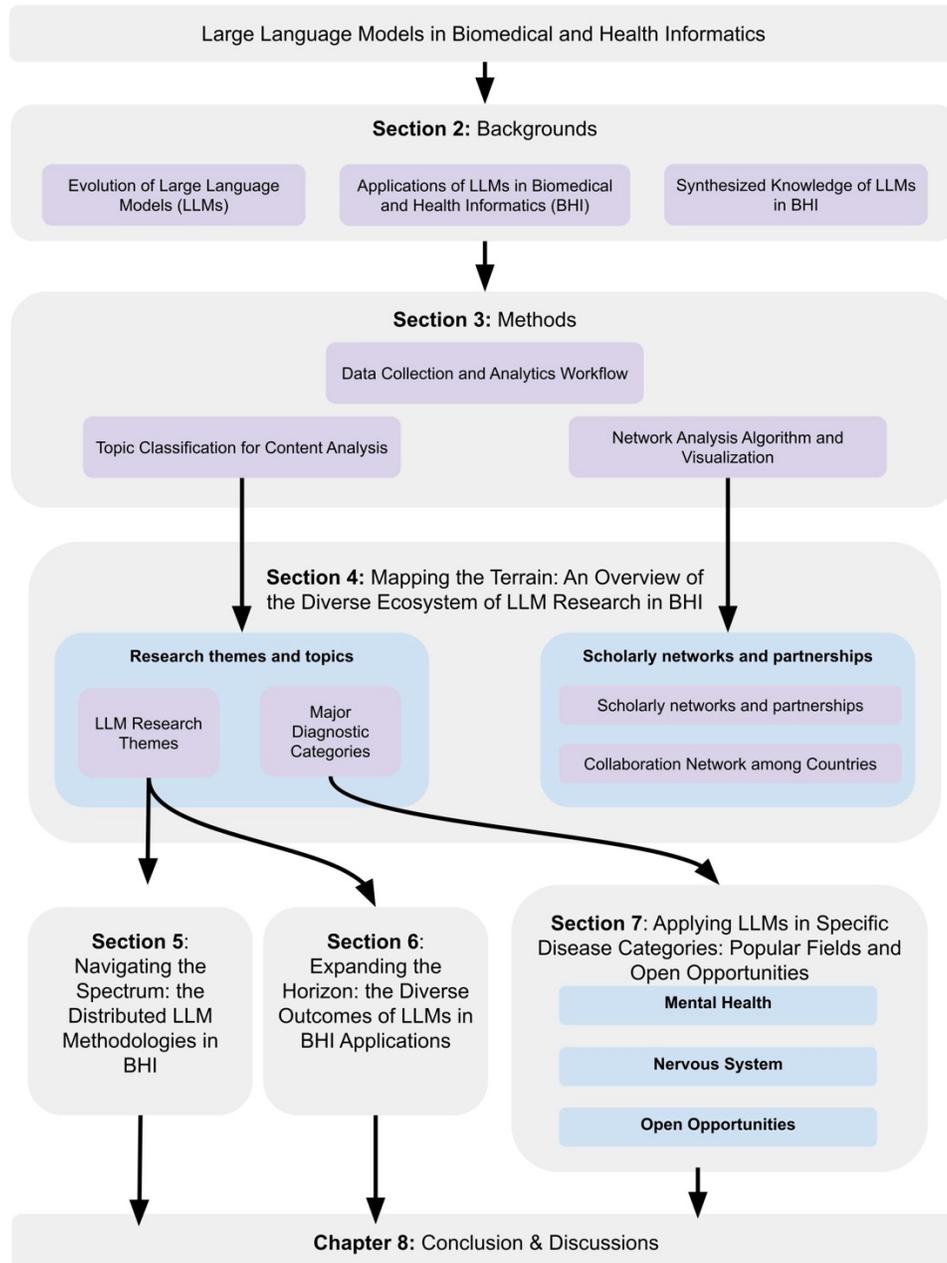

# **Appendix II**. Top 50 Concepts in Keyword Co-occurrence Network Ranked by Total Link Strength

**Table 3** shows the top 50 keywords represented in the Keyword Co-occurrence network by total link strength. The total link strength refers to the sum of the link strengths of one keyword over all the other keywords. The greater the frequency of the co-occurrence, the higher the link strength. Occurrence is the number of times a given keyword appears across the corpus.

**Table 3.** Top 50 Concepts in Keyword Co-occurrence Network

| Keyword | Occurrences | Total Link Strength |
|---|---|---|
| computer science | 1394 | 16099 |
| artificial intelligence | 932 | 11246 |
| medicine | 1033 | 10710 |
| psychology | 719 | 8286 |
| political science | 483 | 5850 |
| law | 461 | 5602 |
| natural language processing | 371 | 4845 |
| mathematics | 344 | 4749 |
| data science | 363 | 4606 |
| biology | 344 | 4585 |
| programming language | 334 | 4471 |
| economics | 303 | 4262 |
| health care | 324 | 4239 |
| philosophy | 294 | 3869 |
| medical education | 349 | 3813 |
| engineering | 301 | 3752 |
| machine learning | 285 | 3699 |
| paleontology | 238 | 3309 |
| pathology | 244 | 2945 |
| context (archaeology) | 187 | 2593 |
| world wide web | 208 | 2514 |
| economic growth | 161 | 2273 |
| operating system | 166 | 2242 |
| linguistics | 172 | 2220 |
| mathematical analysis | 145 | 2134 |
| social psychology | 174 | 2118 |
| task (project management) | 136 | 2117 |
| internal medicine | 208 | 2044 |
| generative grammar | 170 | 2027 |
| epistemology | 142 | 1929 |
| physics | 139 | 1906 |
| language model | 133 | 1845 |
| domain (mathematical analysis) | 123 | 1813 |
| computer security | 140 | 1812 |
| geography | 124 | 1741 |
| management | 114 | 1732 |
| psychiatry | 143 | 1693 |
| set (abstract data type) | 111 | 1552 |
| medical physics | 144 | 1540 |
| sociology | 112 | 1492 |
| quantum mechanics | 102 | 1458 |
| information retrieval | 113 | 1448 |
| family medicine | 134 | 1447 |
| knowledge management | 104 | 1439 |
| pure mathematics | 102 | 1422 |
| field (mathematics) | 98 | 1380 |
| test (biology) | 96 | 1347 |
| medline | 112 | 1205 |
| radiology | 103 | 1204 |
| engineering ethics | 103 | 1198 |

# **Appendix III**. Representative Papers for each LLM Task

**Table 4** presents the representative papers for each LLM task and their respective DOI.

| **Table 4.** Representative Papers for each LLM Task | | |
|---|---|---|
| **LLM Task** | **Representative Paper** | **DOI** |
| Model Evaluation | Evaluating large language models on medical evidence summarization | https://doi.org/10.1038/s41746-023-00896-7 |
| | Assessing the Accuracy and Reliability of AI-Generated Medical Responses: An Evaluation of the Chat-GPT Model | https://doi.org/10.21203/rs.3.rs-2566942/v1 |
| | How Large Language Models Perform on the United States Medical Licensing Examination: A Systematic Review | https://doi.org/10.1101/2023.09.03.23294842 |
| Information Extraction | Exploring zero-shot capability of large language models in inferences from medical oncology notes | https://doi.org/10.48550/arxiv.2308.03853 |
| | Potential of ChatGPT and GPT-4 for Data Mining of Free-Text CT Reports on Lung Cancer | https://doi.org/10.1148/radiol.231362 |
| | Development of a privacy preserving large language model for automated data extraction from thyroid cancer pathology reports | https://doi.org/10.1101/2023.11.08.23298252 |
| Dialogue and Interactive Systems | Understanding the Benefits and Challenges of Deploying Conversational AI Leveraging Large Language Models for Public Health Intervention | https://doi.org/10.1145/3544548.3581503 |
| | A Novel AI-based chatbot Application for Personalized Medical Diagnosis and review using Large Language Models | https://doi.org/10.1109/rmkmate59243.2023.10368616 |
| | ChatGPT: a novel AI assistant for healthcare messaging—a commentary on its potential in addressing patient queries and reducing clinician burnout | https://doi.org/10.1136/leader-2023-000844 |
| Multilinguality | Sailing the Seven Seas: A Multinational Comparison of ChatGPT's Performance on Medical Licensing Examinations | https://doi.org/10.1007/s10439-023-03338-3 |
| | Evaluating the Performance of ChatGPT in a Dermatology Specialty Certificate Examination: A Comparative Analysis between English and Korean Language Settings | https://doi.org/10.21203/rs.3.rs-3241164/v1 |
| | Better to Ask in English: Cross-Lingual Evaluation of Large Language Models for Healthcare Queries | https://doi.org/10.48550/arxiv.2310.13132 |
| Text Generation | Comparing scientific abstracts generated by ChatGPT to real abstracts with detectors and blinded human reviewers | https://doi.org/10.1038/s41746-023-00819-6 |
| | Artificial Intelligence Can Generate Fraudulent but Authentic-Looking Scientific Medical Articles: Pandora's Box Has Been Opened | https://doi.org/10.2196/46924 |
| | Automatic Medical Report Generation via Latent Space Conditioning and Transformers | https://doi.org/10.1109/dasc/picom/cbdcom/cy59711.2023.10361320 |
| Education | Performance of ChatGPT on USMLE: Potential for AI-assisted medical education using large language models | https://doi.org/10.1371/journal.pdig.0000198 |
| | The rise of <scp>ChatGPT</scp>: Exploring its potential in medical education | https://doi.org/10.1002/ase.2270 |
| | Large Language Models in Medical Education: Opportunities, Challenges, and Future Directions | https://doi.org/10.2196/48291 |

| Meta Analysis and Literature Review | Opportunities, Challenges, and Future Directions of Generative Artificial Intelligence in Medical Education: Scoping Review | https://doi.org/10.2196/48785 |
|---|---|---|
| | ChatGPT in Healthcare: A Taxonomy and Systematic Review | https://doi.org/10.1101/2023.03.30.23287899 |
| | Chat GPT in Diagnostic Human Pathology: Will It Be Useful to Pathologists? A Preliminary Review with 'Query Session' and Future Perspectives | https://doi.org/10.3390/ai4040051 |
| Ethics | Ethics of large language models in medicine and medical research | https://doi.org/10.1016/s2589-7500(23)00083-3 |
| | A Medical Ethics Framework for Conversational Artificial Intelligence | https://doi.org/10.2196/43068 |
| | Exploring the potential utility of AI large language models for medical ethics: an expert panel evaluation of GPT-4 | https://doi.org/10.1136/jme-2023-109549 |
| Image, Vision, Video and Multimodality | ChatCAD: Interactive Computer-Aided Diagnosis on Medical Image using Large Language Models | https://doi.org/10.48550/arxiv.2302.07257 |
| | Medical image Generative Pre-Trained Transformer (MI-GPT): future direction for precision medicine | https://doi.org/10.1007/s00259-023-06450-7 |
| | GPT-4 and medical image analysis: strengths, weaknesses and future directions | https://doi.org/10.21037/jmai-23-94 |
| Scholarship and Manuscript Writing | Does GPT-3 qualify as a co-author of a scientific paper publishable in peer-review journals according to the ICMJE criteria? A case study | https://doi.org/10.1007/s44163-023-00055-7 |
| | Guiding principles and proposed classification system for the responsible adoption of artificial intelligence in scientific writing in | https://doi.org/10.3389/frai.2023.1283353 |
| | medicine | |
| | Harnessing Large Language Models in Medical Research and Scientific Writing: A Closer Look to The Future | https://doi.org/10.59707/hymrfbya5348 |
| Inference | Pushing Back on AI: A Dialogue with ChatGPT on Causal Inference in Epidemiology | https://doi.org/10.1007/978-3-031-32013-2_13 |
| | Saama AI Research at SemEval-2023 Task 7: Exploring the Capabilities of Flan-T5 for Multi-evidence Natural Language Inference in Clinical Trial Data | https://doi.org/10.18653/v1/2023.semeval-1.137 |
| | GPT4MIA: Utilizing Generative Pre-trained Transformer (GPT-3) as a Plug-and-Play Transductive Model for Medical Image Analysis | https://doi.org/10.1007/978-3-031-47401-9_15 |
| Summarization | Evaluating Large Language Models on Medical Evidence Summarization | https://doi.org/10.1101/2023.04.22.23288967 |
| | Performance Analysis of Large Language Models for Medical Text Summarization | https://doi.org/10.31219/osf.io/kn5f2 |
| | SummQA at MEDIQA-Chat 2023: In-Context Learning with GPT-4 for Medical Summarization | https://doi.org/10.18653/v1/2023.clinicalnlp-1.51 |
| Sentiment Analysis | Sentiment Analysis of COVID-19 Survey Data: A Comparison of ChatGPT and Fine-tuned OPT Against Widely Used Sentiment Analysis Tools (Preprint) | https://doi.org/10.2196/preprints.50150 |
| | Screening for Depression Using Natural Language Processing (NLP): A Literature Review (Preprint) | https://doi.org/10.2196/preprints.55067 |
| | Applying BERT and ChatGPT for Sentiment Analysis of Lyme Disease in Scientific Literature | https://doi.org/10.48550/arxiv.2302.06474 |

| Named Entity Recognition | DeID-GPT: Zero-shot Medical Text De-Identification by GPT-4 | https://doi.org/10.48550/arxiv.2303.11032 |
| --- | --- | --- |
| | RIGA at SemEval-2023 Task 2: NER Enhanced with GPT-3 | https://doi.org/10.18653/v1/2023.semeval-1.45 |
| | Identification of Ancient Chinese Medical Prescriptions and Case Data Analysis under Artificial Intelligence GPT Algorithm: A Case Study of Song Dynasty Medical Literature | https://doi.org/10.1109/access.2023.3330212 |

# Appendix IV: Analysis of LLM Task: Meta-Analysis and Literature Review

Systematic reviews and meta-analyses in this domain critically assess LLMs, focusing on their capacity to revolutionize various aspects of medical practice [9,226–229] and provide guidelines on their applications [230]. One mainstream in this sub-topic focused on the comprehensive evaluations of different model performances, highlighting the strengths of LLMs in processing medical information and their potential to augment clinical decision-making, while also acknowledging their limitations, such as occasional inaccuracies and biases [231–234]. Detailed investigations into the methodologies reveal how advanced techniques like generative pre-trained transformers [235] and fine-tuning [236] on medical datasets are applied to create innovative applications, from automated medical reporting to virtual patient engagement tools [237]. The other literature suggests future developments, such as emphasizing the need for richer training data [238,239], enhancing interdisciplinary research collaborations [240], and setting up stringent ethical standards to ensure that LLMs can be safely integrated into patient care [231,241]. But they ultimately pave the way for more personalized and efficient healthcare solutions. This collective body of work benchmarks current LLM capabilities and charts a strategic course for their evolution in the healthcare domain.

# Appendix V. Specialized and Contextualized Model Evaluation in Disease Categories

Model Evaluation represents the largest portion of LLM tasks. Specifically, LLMs have been evaluated in their applications for detecting various diseases, from mental health conditions to infectious diseases (**Figure 9** and **Table 5**). The classification tasks are usually the focus of model evaluation.

**Figure 9**: Sankey Diagram of LLM Tasks and Disease Categories (with Paper Count more than 10)

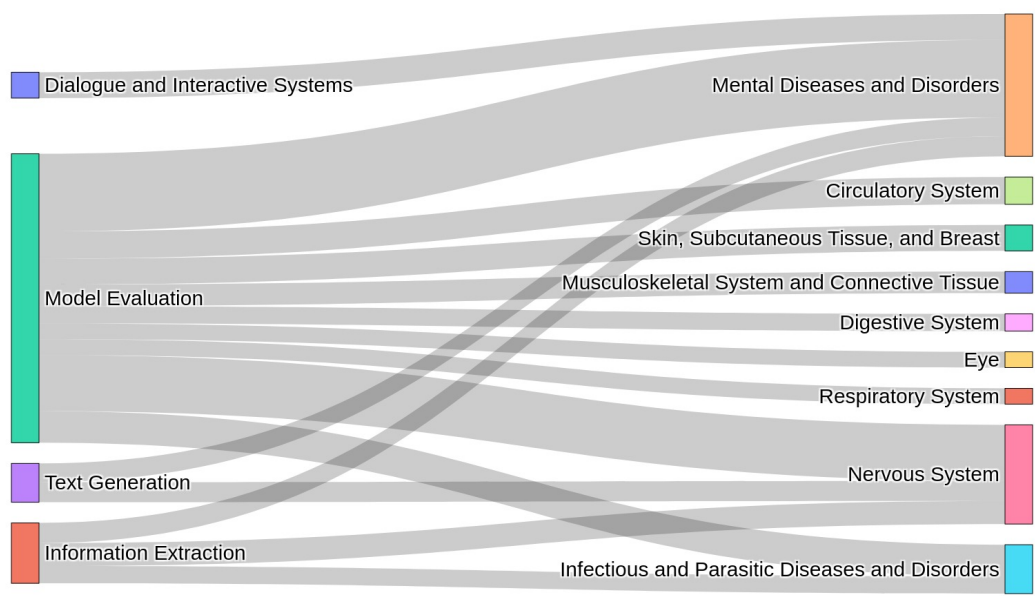

Technical literature on the use of LLMs for mental health analysis has examined the performance of LLMs and LLM-based ChatGPT on basic psychopharmacologic tasks [242], explanation generation of analysis results [243], detection of mental diseases and disorders [244], and so on. Such studies usually evaluate the performance of trained LLMs on pre-labeled datasets compared to a baseline model, with a focus on the accuracy of classification tasks and automatic evaluation metrics [140,244,245]. For instance, [244] evaluates LLM-based ChatGPT on mental health classification tasks with three publically available datasets on stress, depression, and suicidality consisting of annotated social media posts with varying numbers of classes. The model achieved higher classification accuracy compared to a baseline model that always predicted the dominant class.

When datasets are not publically available, researchers come up with classification tasks on their own in specific scenarios [242,246]. For example, [242] created brief vignettes about the decision of selecting antidepressant treatment for adults with major depressive disorder, a basic psychopharmacologic task for clinicians. The authors created and validated the vignettes with experienced clinicians, against which the ChatGPT's ratings of the treatment options are compared.

Explanations of decisions are taken into account in understanding the decisions made by LLMs on classification tasks and analysis of health conditions, and their explainability [209,242,243]. In addition to popular automatic evaluation metrics like perplexity, BLEU-n, and ROUGE-1 [209,245], studies also use human annotation for evaluation and for benchmarking automatic evaluation metrics [243,247]. Additionally, approaches based on prompt engineering are also taken to evaluate the interaction between LLMs and agents by analyzing their mental health referral patterns [248]. Apart from technical literature, other research has also examined and identified the benefits and harms of using LLMs for mental health counseling [249,250] and the issues of hallucination [247].

| **Table 5**: Disease Categories, Paper Counts, and Representative Publications in Model Evaluation Research Category (Paper Count > 10) | | | |
|---|---|---|---|
| **Disease Category** | **Paper Count** | **Example Paper** | **Doi** |
| Mental Diseases and Disorders | 54 | Research Letter: Application of GPT-4 to select next-step antidepressant treatment in major depression | https://doi.org/10.1101/2023.04.14.23288595 |
| | | Benefits and Harms of Large Language Models in Digital Mental Health | https://doi.org/10.48550/arxiv.2311.14693 |
| Nervous System | 39 | Predicting seizure recurrence from medical records using large language models | https://doi.org/10.1016/s2589-7500(23)00205-4 |
| | | The utility of ChatGPT in the assessment of literature on the prevention of migraine: an observational, qualitative study | https://doi.org/10.3389/fneur.2023.1225223 |
| Infectious and Parasitic Diseases and Disorders | 22 | Working With AI to Persuade: Examining a Large Language Model's Ability to Generate Pro-Vaccination Messages | https://doi.org/10.1145/3579592 |

| | | Leveraging Large Language Models and Weak Supervision for Social Media Data Annotation: An Evaluation Using COVID-19 Self-reported Vaccination Tweets | https://doi.org/10.1007/978-3-031-48044-7_26 |
|---|---|---|---|
| Circulatory System | 19 | Uncovering Language Disparity of ChatGPT in Healthcare: Non-English Clinical Environment for Retinal Vascular Disease Classification (Preprint) | https://doi.org/10.2196/preprints.51926 |
| | | ChatGPT Exhibits Gender and Racial Biases in Acute Coronary Syndrome Management | https://doi.org/10.1101/2023.11.14.23298525 |
| Skin, Subcutaneous Tissue, and Breast | 18 | Performance of Three Large Language Models on Dermatology Board Examinations | https://doi.org/10.1016/j.jid.2023.06.208 |
| | | The chatbots are coming: Risks and benefits of consumer-facing artificial intelligence in clinical dermatology | https://doi.org/10.1016/j.jaad.2023.05.088 |
| Musculoskeletal System and Connective Tissue | 15 | Search for Medical Information and Treatment Options for Musculoskeletal Disorders through an Artificial Intelligence | https://doi.org/10.1101/2022.12.16.22283512 |

| | | Chatbot: Focusing on Shoulder Impingement Syndrome | |
| --- | --- | --- | --- |
| | | Large language models and the future of rheumatology: assessing impact and emerging opportunities | https://doi.org/10.1097/bor.0000000000000981 |
| Digestive System | 12 | Advanced prompting as a catalyst: Empowering large language models in the management of gastrointestinal cancers | https://doi.org/10.59717/j.xinn-med.2023.100019 |
| | | Large Language Models for Granularized Barrett's Esophagus Diagnosis Classification | https://doi.org/10.48550/arxiv.2308.08660 |
| Respiratory System | 11 | Natural Language Processing for Covid-19 Consulting System | https://doi.org/10.1016/j.procs.2023.01.112 |
| Eye | 11 | Chat Generative Pretrained Transformer to optimize accessibility for cataract surgery postoperative management | https://doi.org/10.4103/pajo.pajo_51_23 |
| | | Ophtha-LLaMA2: A Large Language Model for Ophthalmology | https://doi.org/10.48550/arxiv.2312.04906 |